# Evolution of interlayer and intralayer magnetism in three atomically thin chromium trihalides

Hyun Ho Kim[1], Bowen Yang[1], Siwen Li[2], Shengwei Jiang[3], Chenhao Jin[3], Zui Tao[3], George Nichols[1], Francois Sfigakis[1], Shazhou Zhong[1], Chenghe Li[4], Shangjie Tian[4], David G. Cory[1], Guo-Xing Miao[1], Jie Shan[3], Kin Fai Mak[3], Hechang Lei[4], Kai Sun[2], Liuyan Zhao[2], and Adam W. Tsen[1*]

[1]*Institute for Quantum Computing, Department of Chemistry, Department of Physics and Astronomy, and Department of Electrical and Computer Engineering, University of Waterloo, Waterloo, Ontario N2L 3G1, Canada*

[2]*Department of Physics, University of Michigan, 450 Church Street, Ann Arbor, Michigan 48109, USA*

[3]*School of Applied and Engineering Physics, Department of Physics, and Kavli Institute at Cornell for Nanoscale Science, Ithaca, NY 14853, USA*

[4]*Department of Physics and Beijing Key Laboratory of Opto-electronic Functional Materials & Micro-Nano Devices, Renmin University of China, Beijing 100872, China*

[*]Correspondence to: awtsen@uwaterloo.ca

## Abstract

We conduct a comprehensive study of three different magnetic semiconductors, $CrI_3$, $CrBr_3$, and $CrCl_3$, by incorporating both few- and bi-layer samples in van der Waals tunnel junctions. We find that the interlayer magnetic ordering, exchange gap, magnetic anisotropy, as well as magnon excitations evolve systematically with changing halogen atom. By fitting to a spin wave theory that accounts for nearest neighbor exchange interactions, we are able to further determine a simple spin Hamiltonian describing all three systems. These results extend the 2D magnetism platform to Ising, Heisenberg, and XY spin classes in a single material family. Using magneto-optical measurements, we additionally demonstrate that ferromagnetism can be stabilized down to monolayer in more isotropic $CrBr_3$, with transition temperature still close to that of the bulk.

The recent discoveries of magnetism in the monolayer limit have opened a new avenue for two-dimensional (2D) materials research[1-4]. Already, several groups have reported a giant tunnel magnetoresistance effect across ultrathin $CrI_3$ layers[5-8] as well as electric field control of their magnetic properties[9-14]. As with $CrI_3$, the entire family of magnetic chromium trihalides ($CrX_3$, X = Cl, Br, and I) possess a layered structure together with relatively strong (weak) in-plane (out-of-plane) exchange coupling[15-20], prompting a thorough investigation of the interlayer and intralayer magnetic properties of all three materials in the two-dimensional (2D) limit.

Within the layers, all three bulk compounds exhibit ferromagnetic (FM) order, although the easy axis is out-of-plane for $CrI_3$ and $CrBr_3$ and in-plane for $CrCl_3$. Interlayer magnetic interactions are not negligible, however, as $CrI_3$[21] and $CrBr_3$[22] are expected to exhibit FM ordering between the layers, while $CrCl_3$[23] shows interlayer antiferromagnetic (AFM) order in the ground state. Yet, in ultrathin $CrI_3$ samples, spins in adjacent layers are instead AFM coupled, giving rise to giant tunnel magnetoresistance when all layers become unipolarized by a relatively small magnetic field[5-8]. Due to the extreme sensitivity of tunnel magnetoresistance to interlayer magnetic order[5-8, 24, 25], we have fabricated graphite/$CrX_3$/graphite tunnel junctions that are fully encapsulated by hexagonal boron nitride (hBN). A schematic illustration of our devices is shown in Fig. 1a and the detailed fabrication procedure can be found in the Methods section. In brief, we exfoliated $CrX_3$ within a nitrogen-filled glovebox and stacked them between top and bottom graphite electrodes before encapsulation by hBN on both sides. Optical images of the devices are shown in SI Appendix, Fig. S1 and their current-voltage characteristics are shown in SI Appendix, Fig. S2.

We begin with temperature-dependent transport behavior under zero magnetic field. In Fig. 1b, we show junction resistance vs temperature upon cooling for three representative devices incorporating the three different trihalides. Their thicknesses measured by atomic force microscopy are $CrI_3$: 5.6 nm, $CrBr_3$: 5.2 nm, and $CrCl_3$: 9 nm, respectively. For easy comparison, the resistances have been normalized by their minimum and maximum values and range between 0 and 1. A marked kink is observed in all devices ($CrI_3$: 46K, $CrBr_3$: 37K, and $CrCl_3$: 17K), close to their respective bulk magnetic transition temperatures ($CrI_3$: 61K[21], $CrBr_3$: 37K[22], and $CrCl_3$: 17K[23]). For magnetic tunnel barriers, it has been found that the resistance either decreases or increases abruptly below the critical temperature depending on whether the magnetic ordering is FM or AFM, respectively[24-26]. This is caused by a spin filtering effect[24, 27], which effectively lowers (raises) the spin-dependent tunnel barrier upon exchange splitting in the FM (AFM) state. A schematic of this effect is shown in the inset of Fig. 1b. Our devices consist of layered magnetic semiconductors in a vertical transport geometry, and therefore we expect our measurements to be most sensitive to the interlayer magnetic ordering of the few-layer samples. We thus assert that $CrCl_3$ and $CrI_3$ exhibit interlayer AFM coupling in their ground state, while $CrBr_3$ shows interlayer FM coupling. For $CrCl_3$ and $CrBr_3$, this is consistent with measurements of the bulk crystal, while those for $CrI_3$ indicate the opposite (FM coupling)[21].

We would like to understand whether the observed interlayer magnetic ordering persists down to the ultimate limit of two atomic layers; however, the resistance kink in the temperature dependence is less apparent for thinner samples due to a smaller spin filtering effect (see SI Appendix, Fig. S3). We therefore turn to the magnetic field dependence. Here, ground state AFM and FM ordering will yield different magnetoresistance behaviors. In Figs. 2a, b, and c, we show resistance vs $B_\perp$ (field perpendicular to the layers) at several different temperatures for the three bilayer (2L) $CrX_3$ devices. In general, the tunneling resistance is smallest when spins in adjacent layers are parallel. First, for 2L $CrI_3$ at low temperature (Fig. 2a), the resistance decreases abruptly when the field exceeds ~0.75T, indicating a spin-flip transition from the AFM ground state (antiparallel out-of-plane) to a parallel spin state at higher field. This resistance change decreases with increasing temperature until it completely disappears above the magnetic transition temperature. These observations are consistent with previous findings[5, 6]. In comparison, the resistance of 2L $CrCl_3$ also decreases substantially with field (Fig. 2c), reflecting that the layers are AFM coupled at zero field. The resistance evolves continuously, however, as spins point in-plane in the ground state and gradually rotate with out-of-plane field. The easy axis of $CrCl_3$ shall be characterized and discussed in more detail later in Figs. 4 and 5. Finally, for 2L $CrBr_3$, the low-temperature resistance is unchanged with field (Fig. 2b), since a spin-parallel FM state has naturally formed and states with both layers spin up or down would show no difference in resistance.

In order to confirm this scenario, we have further performed magnetic circular dichroism (MCD) measurements on another 2L $CrBr_3$ sample (Fig. 3a). Since the MCD signal is proportional to total out-of-plane magnetization, it can resolve the difference between these two spin states with degenerate resistance. The results taken at several different temperatures are shown in SI Appendix, Fig. S4. At low temperature, a finite magnetization is observed at zero field with hysteresis between field sweep up or down, corresponding to switching of the total spin direction of the FM coupled layers. In contrast, bilayer $CrI_3$ shows no net magnetization at zero field as the layers are AFM coupled[1, 5, 9-11]. The critical coercive field needed to flip the spin polarization is also much smaller for $CrBr_3$ (10mT at 5K). We have further performed MCD measurements on 1L, 3L, and 6L $CrBr_3$ and observe similar behavior (Fig 3a and SI Appendix, Fig. S4). The temperature at which the hysteresis disappears is estimated to be 27, 36, and 37K for 1L, 2L, and 3L, respectively. Interestingly, this transition temperature is not much decreased down to monolayer (Fig. 3b).

In addition to interlayer magnetic coupling, we would also like to understand the in-plane magnetic anisotropy of all three 2D compounds in greater detail. We begin with comparing the difference in magnetoresistance behaviors between perpendicular and parallel field configurations for the few-layer devices at low temperature (Figs. 4a and 4c). For CrI$_3$, the critical field needed to fully polarize all the spins in-plane is three times larger than that out-of-plane ($B_\parallel^c = \sim6.5\text{T} \gg B_\perp^c = \sim2\text{T}$). In contrast, the out-of-plane critical field is slightly larger in CrCl$_3$ ($B_\parallel^c = \sim2\text{T} \lesssim B_\perp^c = \sim2.4\text{T}$). For CrBr$_3$, however, magnetic anisotropy cannot be directly determined by magnetoresistance since interlayer FM coupling results in nearly constant resistance independent of field orientation (see SI Appendix, Fig. S6). Instead, we compared the MCD response of few-layer CrBr$_3$ for out-of-plane and in-plane field and obtain $B_\parallel^c = \sim0.44\text{T} \gg B_\perp^c = \sim0.004\text{T}$ (Fig. 4b). Additional information about the layer dependence of the critical fields can be found in SI Appendix, Section IV. These results clearly indicate that the magnetic anisotropy changes with changing halogen atom. We have further measured the full angular dependence of the tunneling current at 2T for few-layer CrI$_3$ and CrCl$_3$ (insets in Figs. 4a and 4c). Similar measurements for other magnetic field levels can be found in SI Appendix, Section V. The results show that CrI$_3$ exhibits the behavior of a highly anisotropic, Ising-type spin system with out-of-plane easy axis. A 2T field applied closely perpendicular to the layers fully polarizes the spins to establish a more conductive state, while the same field applied in-plane only slightly cants the spins to establish a small parallel component. While the easy axis of CrBr$_3$ is also out of plane, the system shows reduced anisotropy in comparison and is closer to Heisenberg. Finally, the easy axis of CrCl$_3$ is in-plane with small anisotropy – it requires a slightly smaller field to rotate the spins within the plane than it does to fully cant them perpendicular, which suggests a weak XY spin model.

These observed differences motivate a detailed microscopic understanding of the spin

Hamiltonian for all three 2D systems, which can be extracted through observation of their excitations (magnons) at low junction biases. Towards this end, we have measured the a.c. conductance (*dI/dV*) vs dc voltage *V* of all three 2L devices using standard lock-in methods (see SI Appendix, Fig. S10). The conductance abruptly increases when the voltage reaches a magnon energy due to the opening of an additional inelastic scattering channel[6, 28, 29]. The magnon energies can then be seen as peaks in the |$d^2I/dV^2$|spectrum. In Fig. 5a, we show, as a color plot, the evolution of |$d^2I/dV^2$| vs *V* with magnetic field along the hard axis for all three 2L trihalides, while similar data along the easy axis is shown in SI Appendix, Fig. S11a. In each case, at least two magnon modes can be seen dispersing with field. This is consistent with the underlying honeycomb lattice, which gives rise two magnon energy branches in momentum space[17]. The magnon density is largest at the M point. The observation of additional peaks indicates that we are resolving magnons with different momenta.

The observed magnon energies can be largely understood by considering only the intralayer magnetic interaction within a single layer. To estimate the effect of interlayer coupling, we note that the easy axis critical field for both $CrI_3$ and $CrCl_3$ (~2T for few layer) decreases substantially with reduced thickness (see SI Appendix, Fig. S5). In particular, it is ~0.1T for 1L $CrI_3$[1]. This indicates that 2T (or 0.2meV for *g* factor = 2) is the energy required to overcome the interlayer AFM coupling for these materials. In contrast, $B_\perp^c$ maintains a small and nearly thickness independent value for $CrBr_3$, which shows interlayer FM coupling. This energy scale is an order of magnitude smaller than the observed magnon energies, and so interlayer coupling should only play a perturbative role.

The minimal model to describe ferromagnetism in a single layer of $CrX_3$ is the 2D anisotropic Heisenberg model, described by the Hamiltonian: $H = -J\sum_{<i,j>}(S_i^x S_j^x + S_i^y S_j^y +$

$\alpha S_i^z S_j^z$), where $S_{i(j)}^{x(y,z)}$ is the spin operator along the $x$ ($y$, $z$) direction at the $Cr^{3+}$ site $i$ ($j$); $J$ is the exchange coupling constant; $\alpha$ is the exchange anisotropy; and $\langle i,j \rangle$ denotes the approximation of the nearest-neighbor exchange coupling. By convention, $z$ is chosen as the direction perpendicular to the layers and $J > 0$ for ferromagnetism. The application of a magnetic field contributes an additional Zeeman term $-g\mu_B B \sum_i S_i$ along the same spin direction.

We have performed a full spin wave analysis for monolayer $CrX_3$ based on the above Hamiltonian on the honeycomb lattice (see SI Appendix, Section VII). The results of which are shown in in Figs. 5b and S11b, and we now summarize. At zero field, the $\Gamma$ and M point magnons have energies $\Gamma_\pm = \frac{9}{2}J(\alpha \pm 1)$ and $M_\pm = \frac{3}{2}J(3\alpha \pm 1)$. For $\alpha$ of order unity, $\Gamma_- \sim 0$ and $M_+ \sim 2M_-$, restricting the magnon assignments in our data. For $CrI_3$ and $CrCl_3$, the most intense peaks are $M_+$ and $M_-$ modes, while the highest energy mode for $CrBr_3$ is assigned to be $\Gamma_+$, although $M_+$ is also faintly visible for positive voltage. We note that for $CrI_3$, this magnon assignment is consistent with a recent neutron scattering study of the bulk crystal[20], which shows comparable magnon energies (~9 and ~15meV) at the M point. At other momenta, it may be important to also consider second- and third-nearest neighbor terms in the spin Hamiltonian.

When the field is applied along the easy axis ($B_\perp$ for $CrI_3$ and $CrBr_3$, and $B_\parallel$ for $CrCl_3$), all magnon energies increase linearly with field with slope $g\mu_B$. We obtain an average $g$ factor of 2.2 between three materials. For field applied in the transverse direction ($B_\parallel$ for $CrI_3$ and $CrBr_3$, and $B_\perp$ for $CrCl_3$), the system undergoes a quantum phase transition as the spins rotate. Here, $\Gamma_+$ and $M_\pm$ modes remain nearly constant up to the anisotropy field, while $\Gamma_-$ gets pushed to zero energy. In Fig. 5a, we indeed observe that the $M_\pm$ peak positions for $CrI_3$ do not shift at low fields. In order to account for the effect of interlayer coupling, we estimate the anisotropy field, $B^a$, for monolayer to be the difference between the critical fields applied along the hard and easy axes for the 2L devices ($B^a$ = 3.63T for CrI$_3$; $B^a$ = 0.44T for CrBr$_3$; and $B^a$ = 0.23T for CrCl$_3$). At high fields, all mode energies again increase by the Zeeman shift. The dashed lines in Fig. 5a and S11a guide the eye to see this change. This simple model captures the essential features of the magnon positions and dispersions for all three compounds, indicating that the data can be largely understood by considering only nearest-neighbor interactions within a single layer.

Importantly, our analysis allows us to extract both the exchange energy $J$ and exchange anisotropy $\alpha$ for the 2D trihalides. In Table 1, we have summarized these values together with other key properties measured in this work. The transition temperature $T_c$, $J$, and $\alpha$ all decrease with smaller halogen atom. We have further measured the low-temperature, exchange gap splitting of the bandstructure $E_{ex}$ in few-layer samples (see SI Appendix, Section VIII), which shows a similar trend. The evolving anisotropy changes the 2D spin class from Ising ($\alpha > 1$) in CrI$_3$, to anisotropic Heisenberg ($\alpha \gtrsim 1$) in CrBr$_3$, and to weak XY ($\alpha \lesssim 1$) in CrCl$_3$. Surprisingly, the transition temperature is not substantially reduced down to 1L for CrBr$_3$ and 2L for CrCl$_3$, despite the low anisotropy in these materials, indicating that strong anisotropy is not necessary to stabilize magnetism in the 2D limit.

We now end by discussing two interesting implications of these results. First, we notice that the transition temperature for bilayer CrBr$_3$ and CrCl$_3$ is already very near that of the bulk crystal, while that for few-layer CrI$_3$ (~46K) is still much reduced from the bulk transition temperature of 61K. It is possible that changing interlayer magnetism from FM to AFM also modifies the transition temperature of this material. In contrast, thin CrBr$_3$ and CrCl$_3$ have similar interlayer coupling with their bulk counterparts. Second, the existence and/or nature of magnetism in monolayer CrCl$_3$ still remains an open question, as the 2D XY model is not expected to show long-range order at finite temperature. It may be possible that interlayer AFM coupling plays a non-negligible role in stabilizing magnetism in bilayers, although one cannot strictly rule out other more complex magnetic orders or the importance of additional in-plane exchange interactions beyond the nearest neighbor. Our work here paves the way for future studies on these topics.

## Methods

**Crystal synthesis.** The single crystals of CrX$_3$ (X = Cl and I) were grown by the chemical vapor transport method. The CrX$_3$ polycrystals were put into the silica tube with the length of 200 mm, and inner diameter of 14mm. The tube was evacuated down to 0.01 Pa and sealed under vacuum. The tubes were placed in two-zone horizontal tube furnace and the source and growth zones were raised to 993 - 873K and 823 - 723K for 24 hours, and then held there for 150 hours. The shiny and plate-like crystals with lateral dimensions up to several millimeters can be obtained. In order to avoid degradation of CrX$_3$ crystals, the samples were stored in glovebox. The CrBr$_3$ single crystals were purchased from HQ graphene.

**Device fabrication.** Graphite (CoorsTek), h-BN (HQ graphene), CrI$_3$, CrBr$_3$ (HQ graphene), and CrCl$_3$ were exfoliated on polydimethylsiloxane(PDMS)-based gel (PF-40/17-X4 from Gel-Pak) within a nitrogen-filled glove box ($P_{O_2}$, $P_{H_2O}$ < 0.1ppm). Pre-patterned Au (40nm)/Ti (5nm) electrodes were fabricated on 285nm-thick SiO$_2$/Si by using conventional photolithography and lift-off methods, and e-beam deposition. Then, vertical heterostructures of hBN/graphite/CrX$_3$/graphite/hBN were sequentially stacked in a home-built transfer setup inside the glove box. The overlapping area of graphite/CrX$_3$/graphite was set to be ~ 10 μm$^2$. 5.6- and 7-

nm-thick $CrI_3$ (8 and 10 layers), 5.2- to 9-nm-thick $CrBr_3$ (8, 10, and 14 layers), and 6- to 9-nm-thick $CrCl_3$ (10, 12, and 15 layers) were used for fabrication. Thin graphite flakes were used as vertical contacts to the $CrX_3$ and connected to the pre-patterned electrodes, while hBN flakes were used as a passivation barrier. Devices were annealed at 393K in the glove box and were stored in vacuum desiccator until the devices were loaded into a cryostat. For bilayer $CrX_3$ devices, sequential pickup[30] was used for fabrication with ~ 1 $\mu m^2$ overlapping area.

**Transport measurements.** Transport measurement was performed in either a He4 cryostat (base temperature 1.4K) or a He3 cryostat (base temperature 0.3K). The dc current/ voltage measurements were performed with a Keithley 2450 source measure unit. A.c. tunneling measurements were performed with an additional lock-in amplifier (Stanford Research Systems SR830 with 100μV a.c. excitation and 77.77Hz frequency). A piezo rotator (atto3DR) was used to rotate the sample relative to the magnetic field.

**Magneto-optical measurements.** The magnetization of hBN-encapsulated $CrBr_3$ flakes was characterized by the magnetic circular dichroism (MCD) microscopy in a He4 cryostat (AttoDry1000) with out-of-plane magnetic field. A diode laser at 405nm with an optical power of 10μW was focused to be a sub-micron spot size on the flakes by an objective of numerical aperture 0.8. The optical excitation was modulated by a photoelastic modulator (PEM) at 50kHz for left and right circular polarization. The laser reflected from $CrBr_3$ was collected by the same objective and then detected by a photodiode.

## Acknowledgements

AWT acknowledges support from an NSERC Discovery grant (RGPIN-2017-03815), an Ontario

Early Researcher Award (ER17-13-199), and the Korea - Canada Cooperation Program through the National Research Foundation of Korea (NRF) funded by the Ministry of Science, ICT and Future Planning (NRF-2017K1A3A1A12073407). G-XM acknowledges support from an NSERC Discovery grant (RGPIN-04178). LZ acknowledges support by NSF CAREER Grant No. DMR-1749774. The magneto-optical measurements at Cornell were supported by NSF (DMR-1807810) and ONR (award N00014-18-1-2368). This research was undertaken thanks in part to funding from the Canada First Research Excellence Fund, the National Key R&D Program of China (No. 2016YFA0300504), and the National Natural Science Foundation of China (Grant No. 11574394, 11774423, and 11822412). We thank Peter Sprenger for the assistance with cryostat operation.

## Author contributions

HHK and AWT designed the experiment. HHK, BY, and SZ fabricated the devices and performed transport measurements with the assistance of GN and FS. CL, ST, and HL synthesized the bulk $CrX_3$ crystals. SL carried out the magnon dispersion calculation for three $CrX_3$ under the guidance of KS and LZ. SJ, CJ, and ZT performed MCD measurements for ultrathin $CrBr_3$ under the guidance of JS and KFM. DGC, G-XM, and AWT supervised. HHK and AWT wrote the paper. All authors provided comments and agreed with the final version of the manuscript.

## Competing interests

The authors declare no competing interests.

## References

1. Huang, B., *et al.* Layer-dependent ferromagnetism in a van der Waals crystal down to the monolayer limit. *Nature*, **546**, 270-273 (2017).

2. Deng, Y., *et al.* Gate-tunable Room-temperature Ferromagnetism in Two-dimensional $Fe_3GeTe_2$. *Nature*, **563**, 94-99 (2018).

3. Gong, C., *et al.* Discovery of intrinsic ferromagnetism in two-dimensional van der Waals crystals. *Nature*, **546**, 265-269 (2017).

4. Bonilla, M., *et al.* Strong room-temperature ferromagnetism in VSe2 monolayers on van der Waals substrates. *Nat. Nanotech.*, **13**, 289-293 (2018).

5. Song, T., *et al.* Giant tunneling magnetoresistance in spin-filter van der Waals heterostructures. *Science*, **360**, 1214-1218 (2018).

6. Klein, D. R., *et al.* Probing magnetism in 2D van der Waals crystalline insulators via electron tunneling. *Science*, **360**, 1218-1222 (2018).

7. Kim, H. H., *et al.* One million percent tunnel magnetoresistance in a magnetic van der Waals heterostructure. *Nano Lett.*, **18**, 4885-4890 (2018).

8. Wang, Z., *et al.* Very Large Tunneling Magnetoresistance in Layered Magnetic Semiconductor $CrI_3$. *Nat. Commun.*, **9**, 2516 (2018).

9. Jiang, S., Shan, J., Mak, K. F. Electric-field switching of two-dimensional van der Waals magnets. *Nat. Mater.*, **17**, 406-410 (2018).

10. Huang, B., *et al.* Electrical control of 2D magnetism in bilayer $CrI_3$. *Nat. Nanotech.*, **13**, 544-548 (2018).

11. Jiang, S., *et al.* Controlling magnetism in 2D $CrI_3$ by electrostatic doping. *Nat. Nanotech.*, **13**, 549-553 (2018).

12. Suarez Morell, E., León, A., Hiroki Miwa, R., Vargas, P. Control of magnetism in bilayer $CrI_3$ by an external electric field. *2D Mater.* **6**, 025020 (2019).

13. Jiang, S., *et al.* Spin tunnel field-effect transistors based on two-dimensional van der Waals heterostructures. *Nat. Electron.* **2**, 159-163 (2019).

14. Song, T., *et al.* Voltage control of a van der Waals spin-filter magnetic tunnel junction. *Nano Lett.* **19**, 915-920 (2019).

15. Liu, J., Sun, Q., Kawazoe, Y., Jena, P. Exfoliating biocompatible ferromagnetic Cr-trihalide monolayers. *Phys. Chem. Chem. Phys.*, **18**, 8777-8784 (2016).

16. Lado, J. L., Fernández-Rossier, J. On the origin of magnetic anisotropy in two dimensional $CrI_3$. *2D Mater.*, **4**, 035002 (2017).

17. Jin, W., *et al.* Raman fingerprint of two terahertz spin wave branches in a two-dimensional honeycomb Ising ferromagnet. *Nat. Commun.*, **9**, 5122 (2018).

18. Zhang, W.-B., Qu, Q., Zhu, P., Lam, C.-H. Robust intrinsic ferromagnetism and half semiconductivity in stable two-dimensional single-layer chromium trihalides. *J. Mater. Chem. C*, **3**, 12457-12468 (2015).

19. Feldkemper, S., Weber, W. Generalized calculation of magnetic coupling constants for Mott-Hubbard insulators: Application to ferromagnetic Cr compounds. *Phys. Rev. B*, **57**, 7755-7766 (1998).

20. Chen, L., *et al.* Topological Spin Excitations in Honeycomb Ferromagnet $CrI_3$. *Phys. Rev. X*, **8**, 041028 (2018).

21. McGuire, M. A., Dixit, H., Cooper, V. R., Sales, B. C. Coupling of Crystal Structure and Magnetism in the Layered, Ferromagnetic Insulator $CrI_3$. *Chem. Mater.*, **27**, 612-620 (2015).

22. Tsubokawa, I. On the Magnetic Properties of a $CrBr_3$ Single Crystal. *J. Phys. Soc. Jpn.*, **15**, 1664-1668 (1960).

23. McGuire, M. A., *et al.* Magnetic behavior and spin-lattice coupling in cleavable van der Waals layered $CrCl_3$ crystals. *Phys. Rev. Mater.*, **1**, 014001 (2017).

24. Hao, X., Moodera, J. S., Meservey, R. Spin-filter effect of ferromagnetic europium sulfide tunnel barriers. *Phys. Rev. B*, **42**, 8235-8243 (1990).

25. Santos, T. S., *et al.* Determining Exchange Splitting in a Magnetic Semiconductor by Spin-Filter Tunneling. *Phys. Rev. Lett.*, **101**, 147201 (2008).

26. Esaki, L., Stiles, P. J., Molnar, S. v. Magnetointernal Field Emission in Junctions of Magnetic Insulators. *Phys. Rev. Lett.*, **19**, 852-854 (1967).

27. Miao, G.-X., Müller, M., Moodera, J. S. Magnetoresistance in Double Spin Filter Tunnel Junctions with Nonmagnetic Electrodes and its Unconventional Bias Dependence. *Phys. Rev. Lett.*, **102**, 076601 (2009).

28. Ghazaryan, D., *et al.* Magnon-assisted tunnelling in van der Waals heterostructures based on $CrBr_3$. *Nat. Electron.*, **1**, 344-349 (2018).

29. Tsui, D. C., Dietz, R. E., Walker, L. R. Multiple Magnon Excitation in NiO by Electron Tunneling. *Phys. Rev. Lett.*, **27**, 1729-1732 (1971).

30. Wang, L., *et al.* One-Dimensional Electrical Contact to a Two-Dimensional Material. *Science*, **342**, 614-617 (2013).

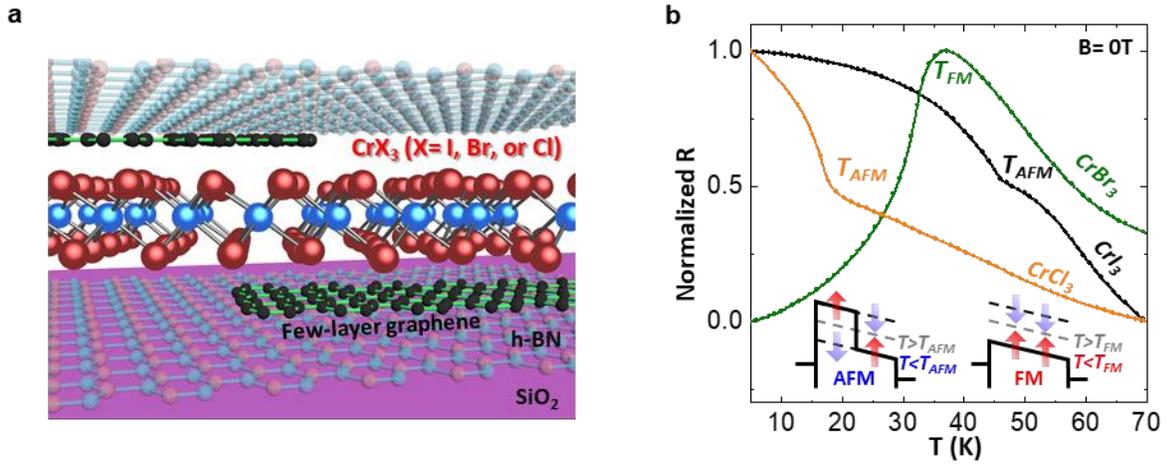

**Figure 1. Magnetic van der Waals tunnel junction incorporating ultrathin chromium trihalides.** (a) Schematic illustration of the device. (b) Normalized temperature-dependent dc resistance of $CrX_3$ (X= I, Br, and Cl) at constant current of 0.1nA. Insets show schematics of the spin-dependent tunnel barrier for AFM and FM interlayer coupling. Red and blue arrows indicate spin orientation and are used throughout.

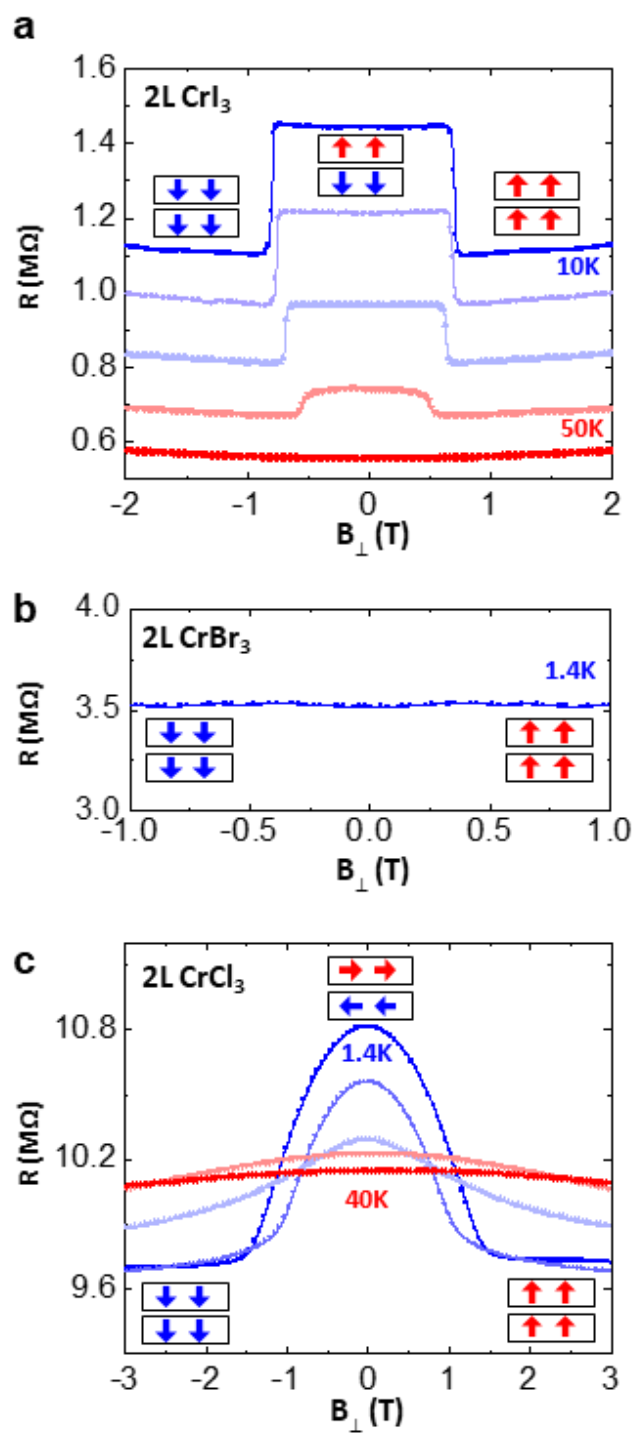

**Figure 2. Tunneling probe of interlayer magnetic coupling in bilayer CrX$_3$.** Resistance vs B$_\perp$ of (a) 2L CrI$_3$ taken at 10, 20, 30, 40, and 50K, in sequence from blue to red (b) 2L CrBr$_3$ at 1.4K, and (c) 2L CrCl$_3$ at 1.4, 10, 20, 30 and 40K. in the sequence from blue to red.

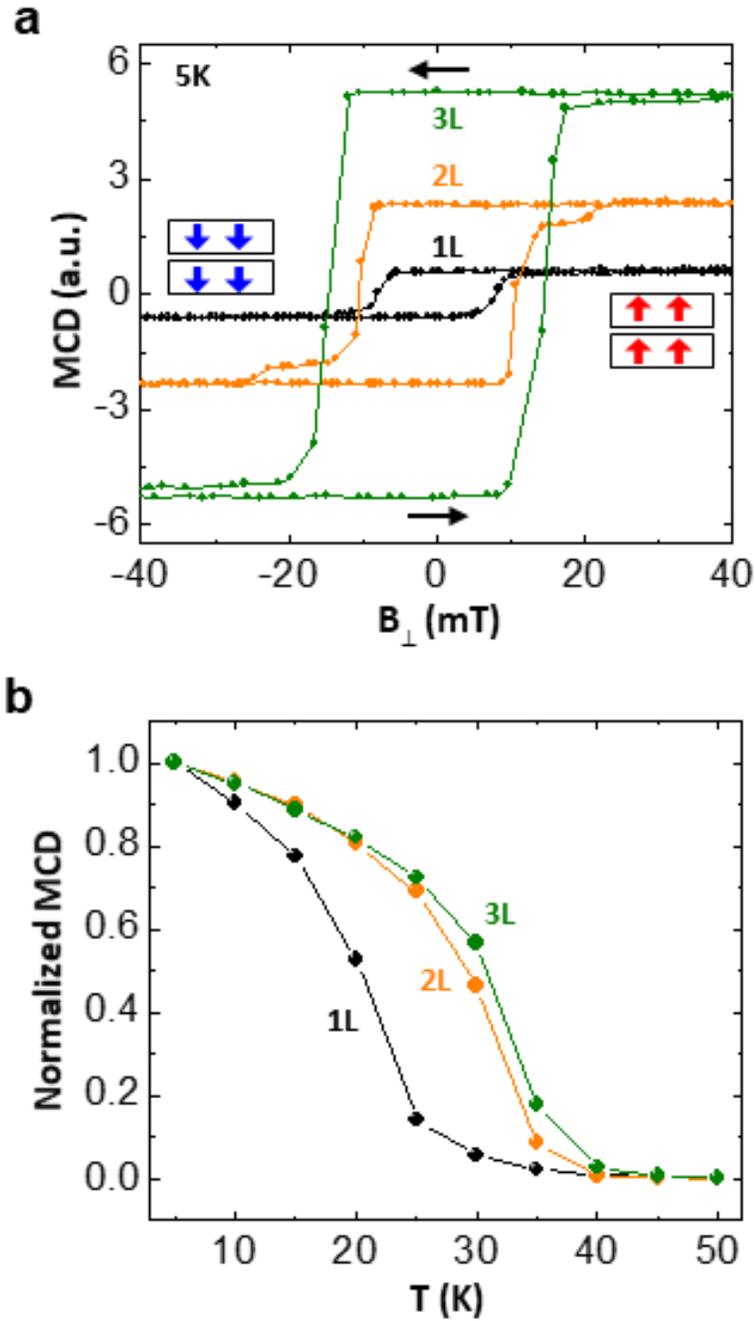

**Figure 3. Magnetic circular dichroism (MCD) measurements on CrBr₃.** (a) Low-temperature MCD vs $B_\perp$ and (b) temperature-dependent normalized MCD at zero field ($\frac{MCD_{\uparrow(\downarrow)}(T)}{MCD_{\uparrow(\downarrow),5K}}$) for 1L, 2L, and 3L CrBr$_3$.

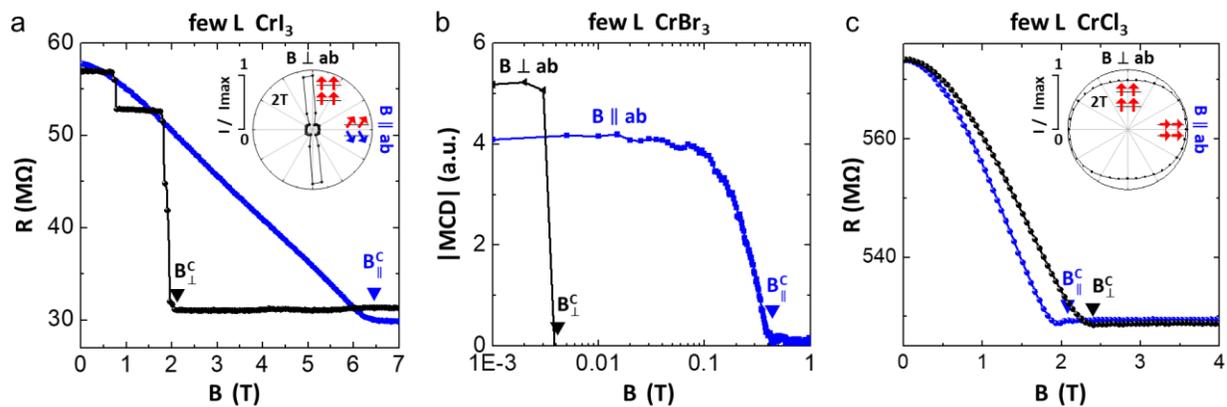

**Figure 4. Magnetic anisotropy in few-layer CrX$_3$.** Comparison of magnetoresistance (1nA current biasing at 1.4K) of (a) 8L CrI$_3$ and (c) 15L CrCl$_3$ for perpendicular and parallel magnetic field directions. (b) |MCD| vs B of 3L CrBr$_3$ at 1.6K for the two field directions. Insets in (a) and (c) show angle-dependent, normalized tunneling current (voltage biasing, 0.5V for CrI$_3$, and 5.7V for CrCl$_3$) at 2T.

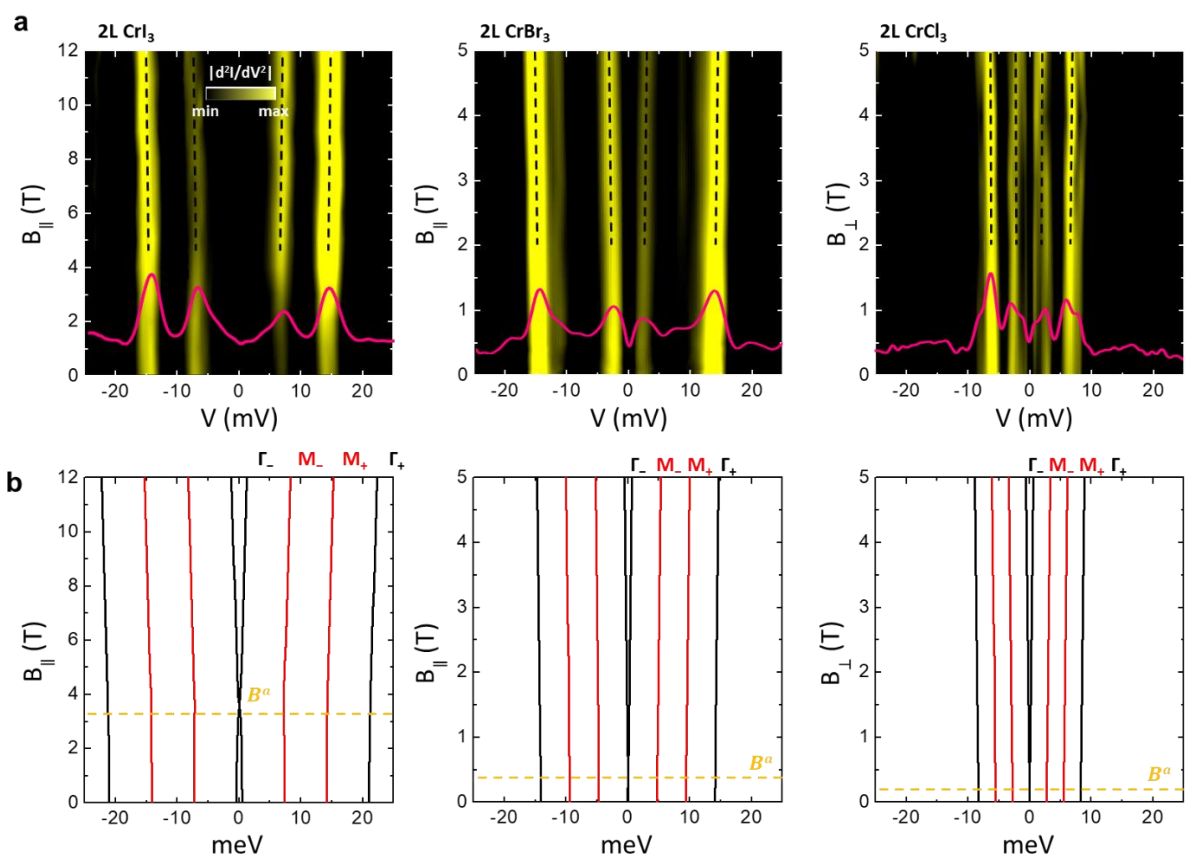

**Figure 5. Inelastic tunneling spectroscopy of magnons in bilayer CrX$_3$.** (a) Field-dependent $|d^2I/dV^2|$ vs voltage for 2L CrX$_3$ at 0.3K and (b) calculated magnon energies for 1L CrX$_3$ with magnetic field applied along the hard axis. Magnon peaks in (a) are partially guided by dashed lines.

**Table 1. Summary of magnetic properties of 2D CrX$_3$**

|  | CrI$_3$ | CrBr$_3$ | CrCl$_3$ |
|---|---|---|---|
| **Interlayer magnetic coupling** | AFM | FM | AFM |
| **T$_C$ (K)** | Few L: 46 (tunneling)<br>2L: 45 (tunneling)<br>1L: 45 (MOKE)[1,11] | Few L: 37 (tunneling)<br>3L: 37 (MCD)<br>2L: 36 (MCD)<br>1L: 27 (MCD) | Few L: 17 (tunneling)<br>2L: 16 (tunneling) |
| **E$_{ex}$ (meV)** | 136 | 122 | 68 |
| **J (meV)** | 2.29 | 1.56 | 0.92 |
| **α** | 1.04 | 1.01 | 0.99 |
| **Spin model** | Ising | Anisotropic Heisenberg | Weak XY |

Supporting Information for

# Evolution of interlayer and intralayer magnetism in three atomically thin chromium trihalides

Hyun Ho Kim[1], Bowen Yang[1], Siwen Li[2], Shengwei Jiang[3], Chenhao Jin[3], Zui Tao[3], George Nichols[1], Francois Sfigakis[1], Shazhou Zhong[1], Chenghe Li[4], Shangjie Tian[4], David G. Cory[1], Guo-Xing Miao[1], Jie Shan[3], Kin Fai Mak[3], Hechang Lei[4], Kai Sun[2], Liuyan Zhao[2], and Adam W. Tsen[1*]

[1]*Institute for Quantum Computing, Department of Chemistry, Department of Physics and Astronomy, and Department of Electrical and Computer Engineering, University of Waterloo, Waterloo, Ontario N2L 3G1, Canada*

[2]*Department of Physics, University of Michigan, 450 Church Street, Ann Arbor, Michigan 48109,USA*

[3]*School of Applied and Engineering Physics, Department of Physics, and Kavli Institute for Nanoscale Science, Cornell University, Ithaca, New York 14853, USA*

[4]*Department of Physics and Beijing Key Laboratory of Opto-electronic Functional Materials & Micro-Nano Devices, Renmin University of China, Beijing 100872, China*

[*]Correspondence to: awtsen@uwaterloo.ca

## I. Fabrication of CrX₃ devices

Optical images of CrX$_3$ devices were taken just before top hBN passivation, as shown in Fig. S1. Cracks and wrinkles were avoided during the fabrication. For few-layer devices, we used the sequential transfer method with PDMS as our previous work[1]. Conventional pickup process using poly(bisphenol A carbonate) thin film[2] was used for the fabrication of bilayer devices due to a cracking issue.

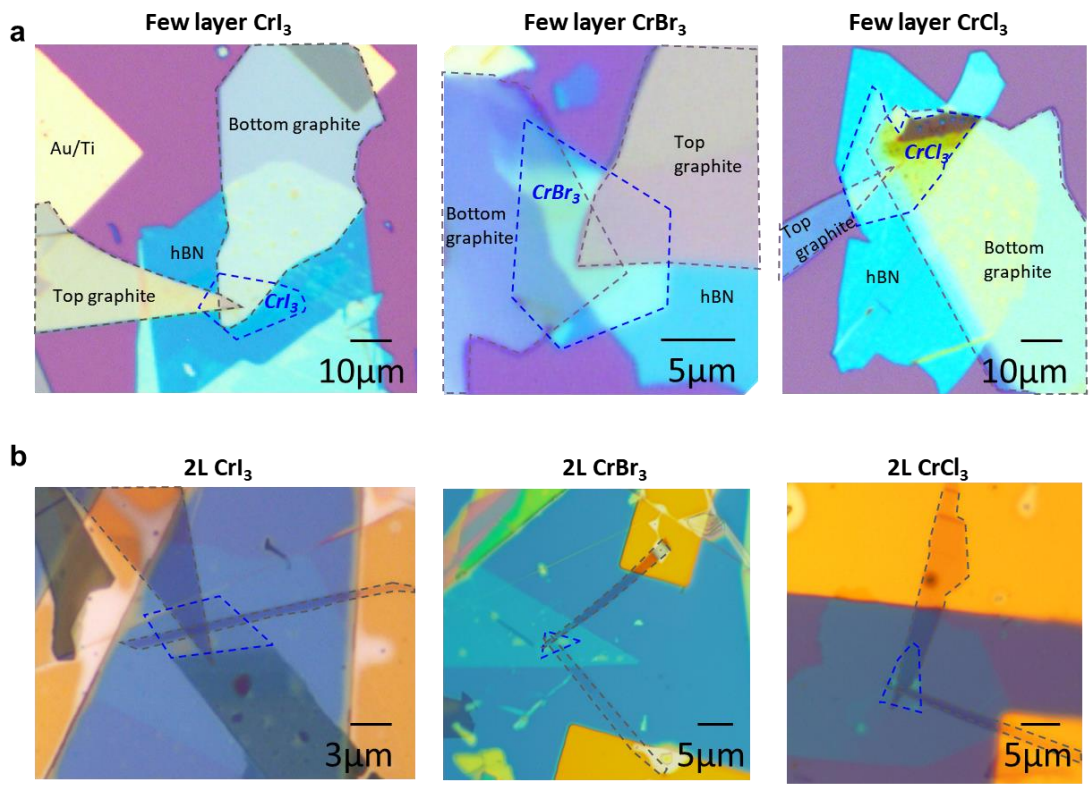

**Figure S1.** Optical images of vertical van der Waals magnetic tunnel junction devices. (a) 8-layer CrI$_3$, 14-layer CrBr$_3$, and 15-layer CrCl$_3$ (before covering top h-BN). (b) complete 2L CrX$_3$ devices.

## II. Additional information for temperature dependence

In Fig. S2a, we show nonlinear current-voltage behavior characteristics of few-layer $CrX_3$ magnetic tunnel junctions. Below the magnetic transition temperature ($T_{AFM}$ and $T_{FM}$), current at fixed voltage decreases as the temperature decreases in $CrI_3$ and $CrCl_3$ devices, whereas $CrBr_3$ shows a reverse tendency. Bilayer devices exhibit near ohmic behavior in $CrI_3$ and $CrBr_3$, indicating that direct tunneling is dominant rather than Fowler-Nordheim (FN) tunneling. Since the bandgap is expected to be large in $CrCl_3$, we could still measure FN tunneling starting from ~1V. Due to this effect, clear spin-filtering can still be observed in 2L $CrCl_3$ devices, but not in 2L $CrI_3$ or 2L $CrBr_3$, as shown in Fig. S3.

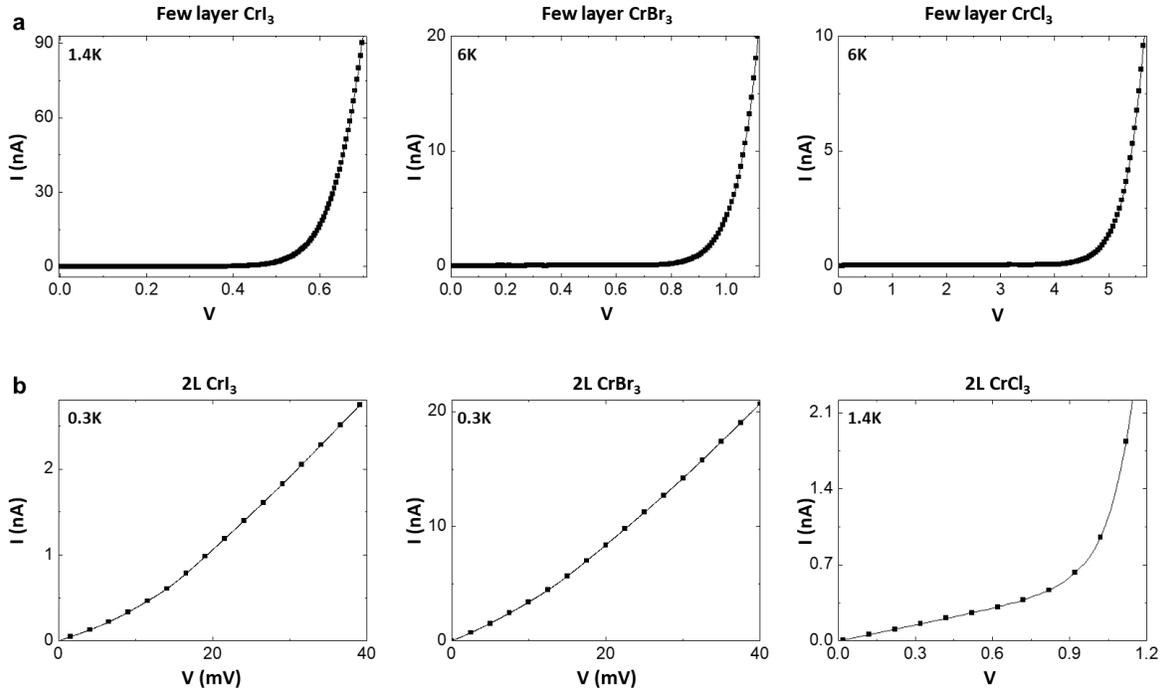

**Figure S2.** Low-temperature I-V measurement of (a) few-layer $CrX_3$ (8L $CrI_3$, 8L $CrBr_3$, and 15L $CrCl_3$. And (b) 2L $CrX_3$.

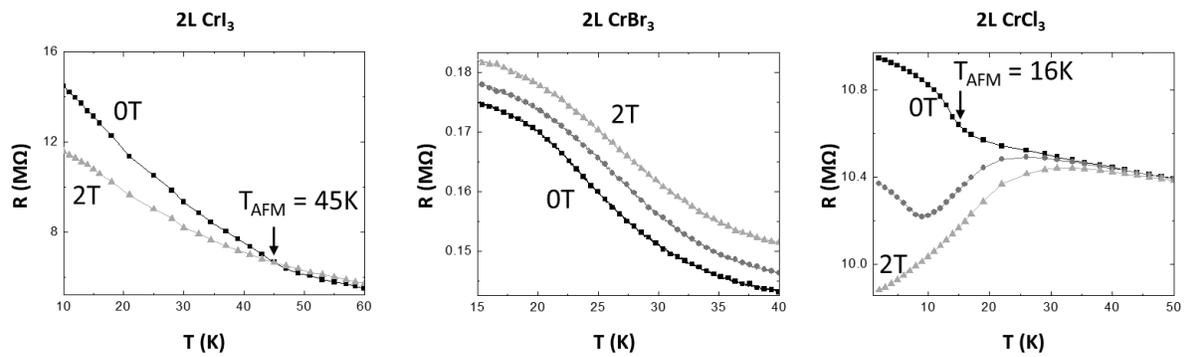

**Figure S3.** Resistance vs temperature as a function of perpendicular field (0, 1, and 2T) for three bilayer $CrX_3$ devices.

## III. Magnetic circular dichroism measurements on CrBr$_3$

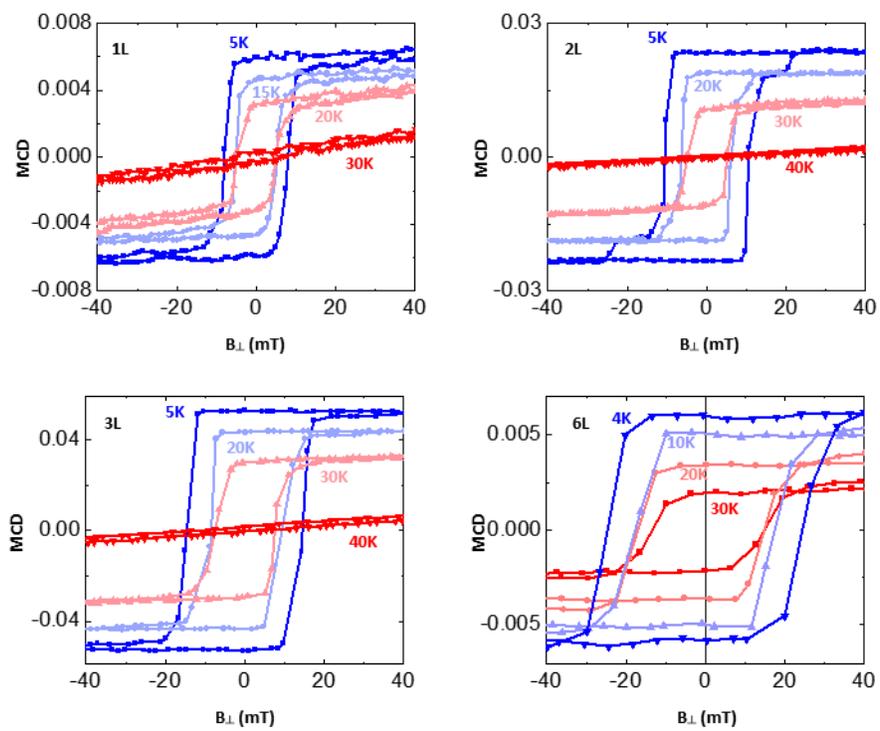

**Figure S4.** Magnetic circular dichroism (MCD vs $B_\perp$) measurements on 1L, 2L, 3L, and 6L CrBr$_3$.

## IV. Layer-dependent critical field of CrX$_3$

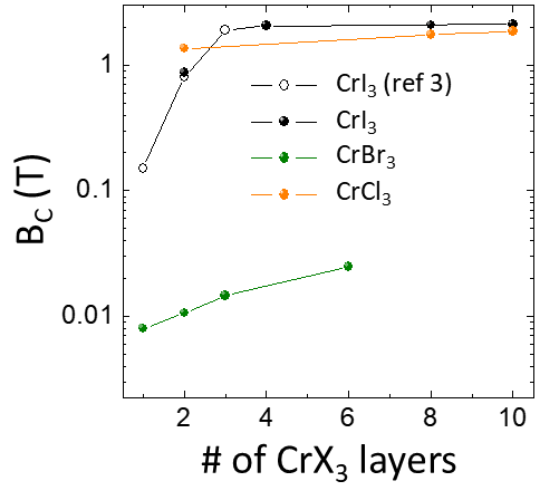

**Figure S5.** Critical field (B$_C$) of CrX$_3$ as a function of the number of layers along with easy axis. CrI$_3$ and CrBr$_3$ were measured with perpendicular field and CrCl$_3$ was measured with parallel field. It is noted that critical fields for CrI$_3$ and CrCl$_3$ above 2L show relatively higher values due to the interlayer AFM coupling.

## V. Additional information for magnetic anisotropy

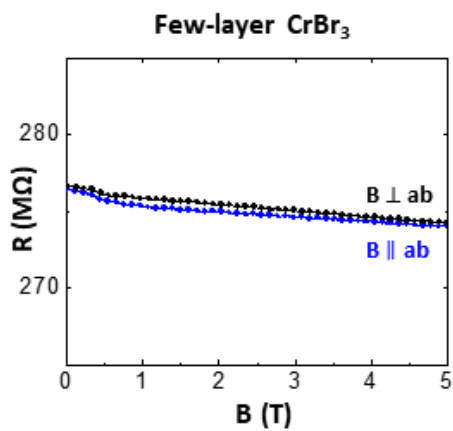

**Figure S6.** Comparison of magnetoresistance (1nA current biasing at 1.4K) of 10L $CrBr_3$.

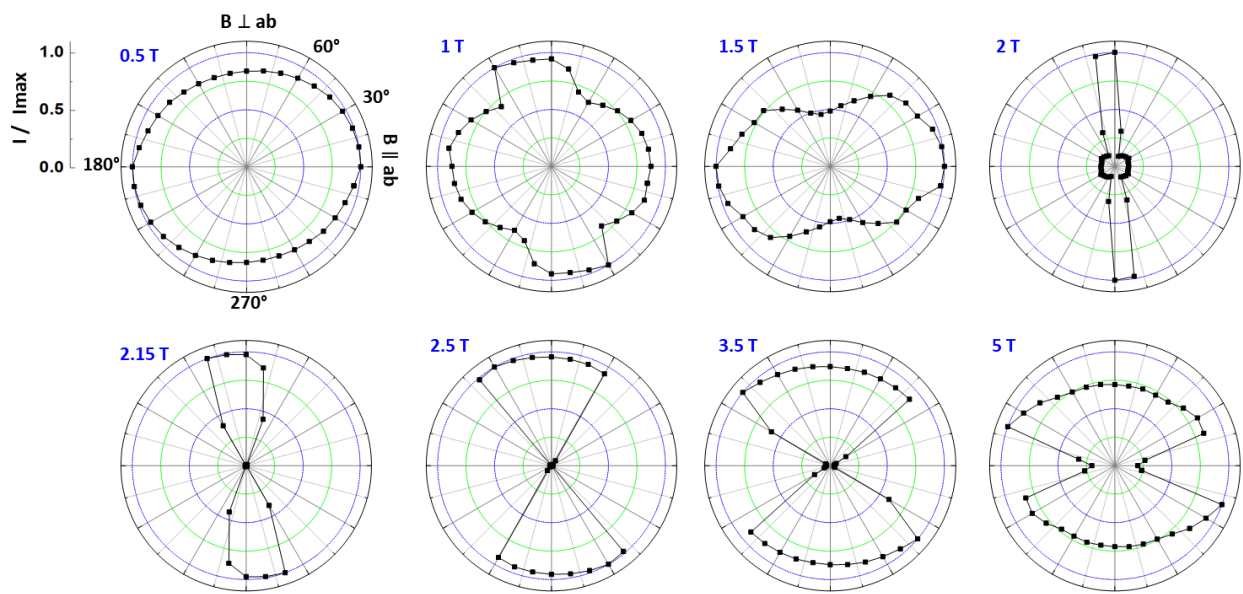

**Figure S7.** Theta-and field-dependent normalized tunnel current of 8-layer $CrI_3$ from 0.5 to 5T.

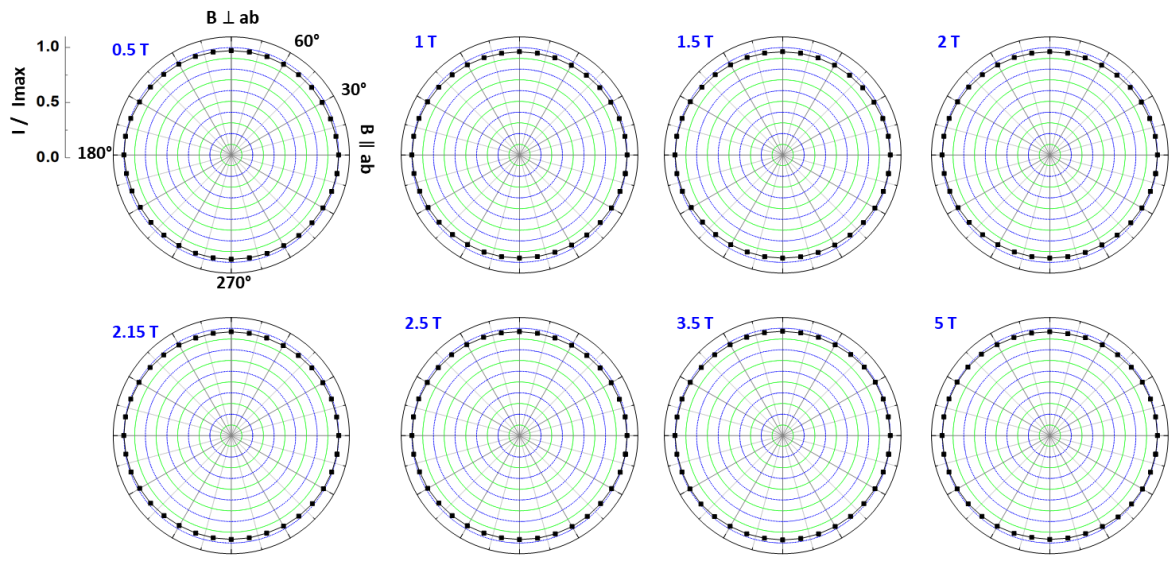

**Figure S8.** Theta-and field-dependent normalized tunnel current of 10-layer CrBr$_3$ from 0.5 to 5T.

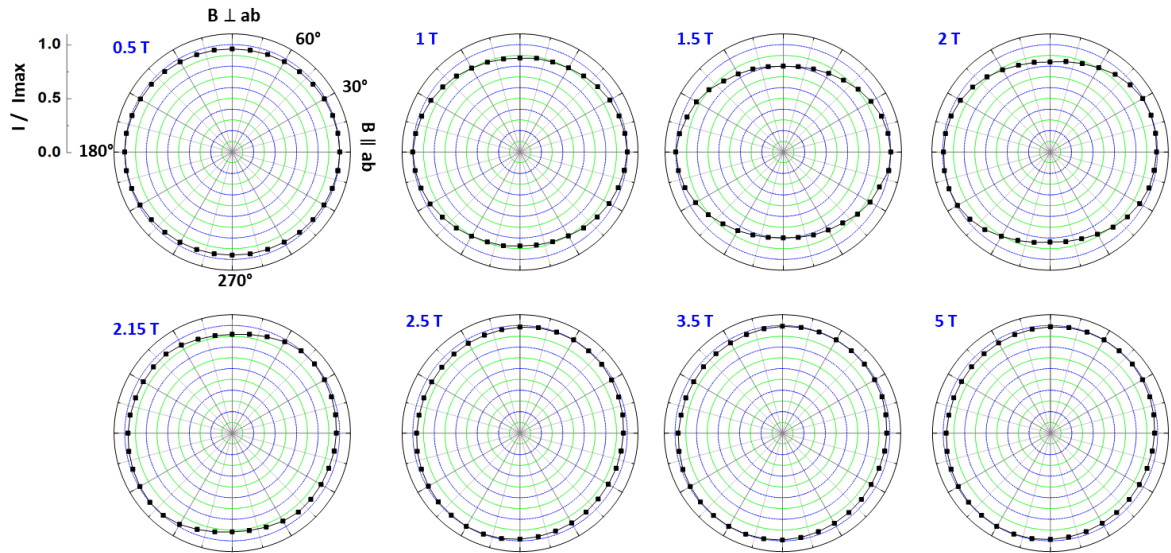

**Figure S9.** Theta-and field-dependent normalized tunnel current of 15-layer CrCl$_3$ from 0.5 to 5T.

## VI. Additional information for inelastic tunneling spectroscopy

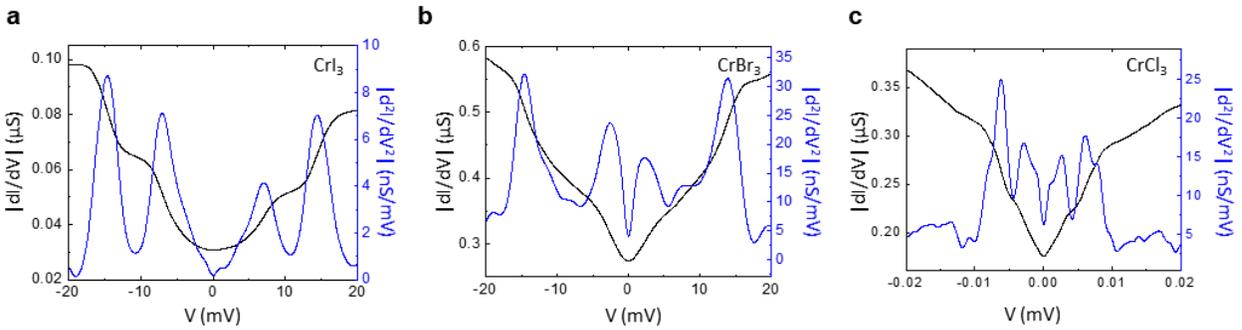

**Figure S10.** The ac conductance (G, $dI/dV$) vs dc voltage of (a) 2L $CrI_3$, (b) 2L $CrBr_3$, and (c) 2L $CrCl_3$. The derivative $|d^2I/dV^2|$ is shown in blue solid line. Conductance at each dc voltage is measured with 100μV ac excitation voltage.

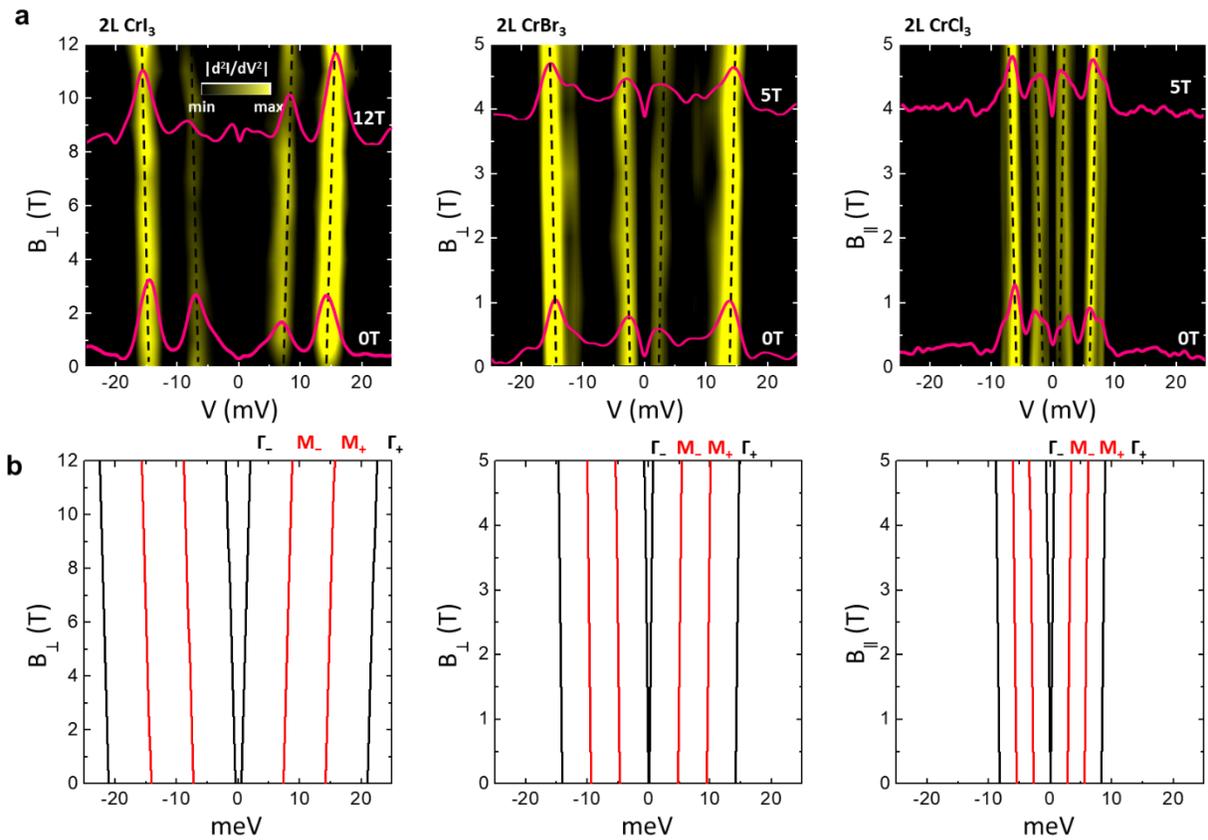

**Figure S11.** (a) Field-dependent $|d^2I/dV^2|$ vs voltage for 2L $CrX_3$ at 0.3K and (b) calculated magnon energies for 1L $CrX_3$ with field applied along the easy axis. Magnon peaks in (a) are guided by dashed lines.

## VII. Spin wave calculations under an external magnetic field

In the following, we perform calculations of spin wave dispersions under an external magnetic field for monolayer honeycomb magnets CrX$_3$. Our spin Hamiltonian only considers monolayer spin structure instead of the actual bilayer structure, because interlayer interactions are small compared to intralayer exchange interactions. Although bilayer splitting modifies the monolayer spin wave dispersions, it seems unnecessary in evaluating intralayer exchange interactions. We choose the anisotropic Heisenberg spin model below,

$$H = -J \sum_{<i,j>} \left( S_i^x S_j^x + S_i^y S_j^y + \alpha S_i^z S_j^z \right) - g\mu_B \left( B_z \sum_i S_i^z + B_x \sum_i S_i^x \right)$$

where $J$ is the nearest neighbor exchange energy, $i$ and $j$ denote the two inequivalent Cr$^{3+}$ site, and $\alpha$ scales the $z$-direction exchange strength. In particular, $\alpha$ is greater than, equivalent to, or smaller than 1 for Ising, isotropic Heisenberg, or XY magnets, respectively. $g$ is the $g$-factor for Cr$^{3+}$ magnetic moments, $\mu_B$ is the Bohr magneton, and $\vec{B}$ is external magnetic field. Depending on the experimental geometry, either $B_z$ or $B_x$ is nonzero, representing the magnetic field perpendicular to or parallel to the plane of CrX$_3$ layers.

Without an external magnetic field, the spins in CrX$_3$ align either along the $z$-axis or in the $xy$-plane depending on the magnitude of $\alpha$. When an external field is applied perpendicular to the easy axis or easy plane, it tilts the spins off its original direction, to a new direction that minimizes the energy for the Hamiltonian above. To keep it general for all three types of magnets, we characterize the spin orientation by its tilt angle from $z$ axis, $\theta$. For example, $\theta = 0$ corresponds to the spins aligning along the $z$-axis and $\theta = \pi/2$ for spins in the $xy$-plane. To determine $\theta$ for a given external magnetic field, a rotation transform is applied, where $\vec{\tilde{S}}$ is the spin orientation in the new ground state under the external field:

$$\begin{cases} S^x = \tilde{S}^x \cos\theta + \tilde{S}^z \sin\theta \\ S^z = -\tilde{S}^x \sin\theta + \tilde{S}^z \cos\theta \end{cases}$$

After applying Holstein-Primakoff transform for both site $i$ and site $j$,

$$\begin{cases} \tilde{S}^z = S - a^\dagger a \\ \tilde{S}^+ = \sqrt{2S}(1 - \frac{a^\dagger a}{2S})a \\ \tilde{S}^- = \sqrt{2S} a^\dagger \end{cases}$$

we arrive at the following Hamiltonian where spin wave interactions are ignored:

$$H = H_{const} + H_1 + H_2$$

$$H_{const} = -SN(JzS(\alpha \cos^2\theta + \sin^2\theta) + 2g\mu_B(\sin\theta B_x + \cos\theta B_z))$$

$$H_1 = (\frac{1}{2}JSz(\alpha - 1)\sin 2\theta + g\mu_B(B_z \sin\theta - B_x \cos\theta))\sqrt{S/2} \sum_i (a_i^\dagger + a_i)$$

$$H_2 = -\frac{1}{2}JS(\alpha \sin^2\theta + \cos^2\theta - 1) \sum_{<i,j>} (a_i^\dagger b_j^\dagger + a_i b_j)$$

$$- \frac{1}{2}JS(\alpha \sin^2\theta + \cos^2\theta + 1) \sum_{<i,j>} (a_i^\dagger b_j + a_i b_j^\dagger)$$

$$+ JS(\alpha \cos^2\theta + \sin^2\theta) \sum_{<i,j>} (a_i^\dagger a_i + b_j^\dagger b_j)$$

$$+ g\mu_B(\sin\theta B_x + \cos\theta B_z)(\sum_i a_i^\dagger a_i + \sum_j b_j^\dagger b_j)$$

where $N$ is the number of sublattices, and $z$ is the number of nearest neighbor spins. Minimizing $H_{const}$ yields the new spin orientation $\theta$ and eliminates $H_1$. If $\vec{B}$ is parallel to the original spin orientation, the spins remain the same. If $\vec{B}$ is perpendicular, $\theta$ is a function of $|\vec{B}|$. At low fields,

$$\theta = \arcsin\frac{g\mu_B B_x}{JSz(\alpha - 1)} \quad (B_z = 0, \alpha > 1)$$

and

$$\theta = \arccos \frac{g\mu_B B_z}{JSz(1-\alpha)} \quad (B_x = 0, \alpha < 1)$$

A critical field strength $B_c$ exists above which the spins are completely aligned along the external field direction.

To obtain the spin wave dispersions under external magnetic fields, Fourier and Bogoliubov transforms of spin wave operators are applied to $H_2$. Fourier transform yields

$$H_2 = Q_1 \sum_k (a_k^\dagger b_{-k}^\dagger \gamma_{-k} + a_k b_{-k} \gamma_k) + Q_2 \sum_k (a_k^\dagger b_k \gamma_{-k} + a_k b_k^\dagger \gamma_k)$$

$$+ Q_3 \sum_k (a_k^\dagger a_k + b_k^\dagger b_k)$$

where

$$\gamma_k = \frac{1}{z} \sum_\delta e^{ik\delta}$$

$$Q_1 = -\frac{1}{2} JSz(\alpha \sin^2 \theta + \cos^2 \theta - 1)$$

$$Q_2 = -\frac{1}{2} JSz(\alpha \sin^2 \theta + \cos^2 \theta + 1)$$

$$Q_3 = JSz(\alpha \cos^2 \theta + \sin^2 \theta) + g\mu_B(\sin\theta B_x + \cos\theta B_z)$$

$a_k$ and $b_k$ correspond to the Fourier transformed spin wave operators of real space operators $a_i$ and $b_j$, respectively. Bogoliubov transform for bosonic excitations is then done by

$$\begin{pmatrix} \phi_{-k} \\ \psi_{-k} \\ \phi_k^\dagger \\ \psi_k^\dagger \end{pmatrix} = M \begin{pmatrix} a_{-k} \\ b_{-k} \\ a_k^\dagger \\ b_k^\dagger \end{pmatrix}$$

$$M = \begin{pmatrix} \dfrac{Q_3 - Q_2 P_1 - P_2}{Q_1 \gamma_k} & -\dfrac{Q_3 - Q_2 P_1 - P_2}{Q_1 P_1} & -\sqrt{\dfrac{\gamma_{-k}}{\gamma_k}} & 1 \\ \dfrac{Q_3 + Q_2 P_1 - P_3}{Q_1 \gamma_k} & \dfrac{Q_3 + Q_2 P_1 - P_3}{Q_1 P_1} & \sqrt{\dfrac{\gamma_{-k}}{\gamma_k}} & 1 \\ \dfrac{Q_3 - Q_2 P_1 + P_2}{Q_1 \gamma_k} & -\dfrac{Q_3 - Q_2 P_1 + P_2}{Q_1 P_1} & -\sqrt{\dfrac{\gamma_{-k}}{\gamma_k}} & 1 \\ \dfrac{Q_3 + Q_2 P_1 - P_3}{Q_1 \gamma_k} & \dfrac{Q_3 + Q_2 P_1 - P_3}{Q_1 P_1} & \sqrt{\dfrac{\gamma_{-k}}{\gamma_k}} & 1 \end{pmatrix}$$

where

$$P_1 = \sqrt{\gamma_k \gamma_{-k}}$$

$$P_2 = \sqrt{Q_3^2 - 2Q_3 Q_2 P_1 + (Q_2^2 - Q_1^2) P_1^2}$$

$$P_3 = \sqrt{Q_3^2 + 2Q_3 Q_2 P_1 + (Q_2^2 - Q_1^2) P_1^2}$$

$M$ is the Bogoliubov transform matrix under which the new operators $\phi_k$ and $\psi_k$ satisfy bosonic commutation relations (up to a normalization factor). Spin wave dispersions as a function of external field strength are obtained as $E_1 = P_2$, and $E_2 = P_3$. If $\vec{B}$ is parallel to the original spin orientation, spin wave energies increase by $g\mu_B |\vec{B}|$. If $\vec{B}$ is perpendicular, spin wave energies show an anomaly at $B_c$, as shown in the main text Figure 5.

**Input parameters and fit with experimental data**

For $Cr^{3+}$, the spin moment has $S = 3/2$. In the honeycomb lattice, the number of nearest neighbors is $z = 3$. At large fields, the spin wave energies increase by $g\mu_B B$. $g$-factor is then extracted from the slope of the spin wave energies as a function of external field. $g$ is taken to be

2.1788, which is the average of the slopes of all three compounds. The spin wave density of states is high at Γ- and M-point, so that the experimentally observed peaks are related with two Γ-point excitations and two M-point excitations. We use the transition fields $B_c$ from the field dependent tunneling spectroscopy data to find a relation between $J$ and $\alpha$, and identify their values fitting the calculated functional forms to the experimental results.

**Table S1** lists $B_c$ from experiment and $J$ and $\alpha$ from fitting results.

**Table S1.** Spin wave parameters in $CrX_3$ compounds

| Compound | $B_c$ (T) | $J$ (meV) | $\alpha$ |
|---|---|---|---|
| $CrI_3$ | 3.63 | 2.286 | 1.0445 |
| $CrBr_3$ | 0.44 | 1.562 | 1.0079 |
| $CrCl_3$ | 0.23 | 0.9208 | 0.9930 |

## VIII. Estimation of spin-splitting energy

In the top panel of Fig. S12a, we show how to deduce the exchange energy gap from the current-voltage characteristics. As described in a previous report[1], the spin-slitting energy gap ($E_{ex}$) can be approximated by taking barrier height difference between fully polarized P state and paramagnetic state (PM), $E_{ex} = 2(\Phi_{PM} - \Phi_P)$. Barrier height is calculated in Fowler-Nordheim[3] regime using the current-voltage relation:

$$J = \left(\frac{e^3 E^2}{4\hbar\Phi}\right) \exp\left(-\frac{4\sqrt{2m\Phi^3}}{3\hbar e E}\right)$$

where $e$ is electronic charge, $E$ is the electric field across the $CrX_3$, $m$ is the effective electron mass, which we estimate to be the free-electron mass, $\hbar$ is the Planck's constant, and $\Phi$ is the barrier height between the thin graphite electrode and $CrX_3$. In the bottom panel of Fig. S12a, we plot representative $\ln(J/E^2)$ vs $E^{-1}$ for a 10-layer $CrI_3$ in both the PM state (at 49K with B=0) and fully polarized state (at 1.4K with $B_\perp = 5.5T$). We found that the negative slope proportional to $\Phi^{3/2}$ becomes flatter from PM to fully parallel state. As shown in Fig. S12b, the splitting energies of few-layer $CrX_3$ are estimated to be $CrI_3$ ~136meV, $CrBr_3$ ~122meV, and $CrCl_3$ ~68meV. While these values are slightly smaller than those from bandstructure calculations[4], the trend with changing halogen size is consistent. At the same time, we also observed different initial barrier heights in PM state (~227meV in $CrI_3$, ~538meV in $CrBr_3$, and ~943meV in $CrCl_3$), which is consistent with calculations showing increasing bandgap with decreasing halogen atom size. Among the three $CrX_3$, $CrI_3$ exhibits the largest exchange splitting gap, which can be explained by a stronger intralayer Cr-Cr superexchange.

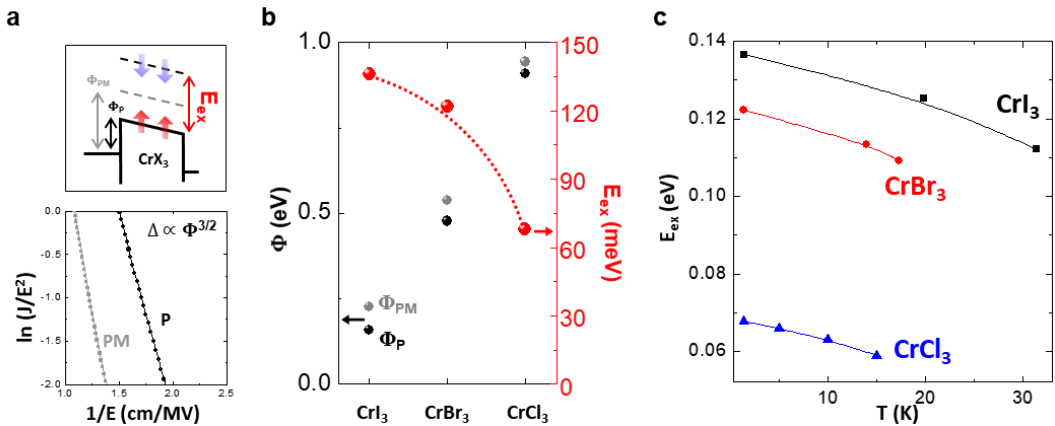

**Figure S12. Magnetic exchange splitting gap of few-layer CrX$_3$.** (a) Top: a schematic illustration of band structure for the calculation of splitting energy gap ($E_{ex} = 2(\Phi_{PM} - \Phi_P)$). Bottom: ln (J/E$^2$) versus E$^{-1}$ plot of 10-layer CrI$_3$ at two different conditions (grey: 49K without field, black: 1.4K with B ⊥ ab = 5.5T). (b) $\Phi_{PM}$ and $\Phi_P$ for 10L CrI$_3$, 10L CrBr$_3$, and 12L CrCl$_3$ and deduced exchange splitting gap at 1.4K.

## References

1. Kim, H. H., *et al.* One million percent tunnel magnetoresistance in a magnetic van der Waals heterostructure. *Nano Lett.*, **18**, 4885-4890 (2018).

2. Wang, L., *et al.* One-Dimensional Electrical Contact to a Two-Dimensional Material. *Science*, **342**, 614-617 (2013).

3. Lenzlinger, M., Snow, E. H. Fowler-Nordheim Tunneling into Thermally Grown SiO$_2$. *J. Appl. Phys.*, **40**, 278-283 (1969).

4. Zhang, W. B., Qu, Q., Zhua, P., Lam, C. H. Robust intrinsic ferromagnetism and half semiconductivity in stable two-dimensional single-layer chromium trihalides. *J. Mater. Chem. C*, **3**, 12457-12468 (2015).